\documentclass[12pt]{article}
\usepackage{amssymb,latexsym,epsfig,cite}

\begin{document}

\title{Asymptotic singular behaviour of inhomogeneous cosmologies in
Einstein-Maxwell-dilaton-axion theory}
\author{ Luis A. L\'opez and
Nora Bret\'on.}


\maketitle

\begin{abstract}
We present the study of exact inhomogeneous cosmological solutions to a
four-dimensional low energy limit of string theory containing non-minimal
interacting electromagnetic, dilaton and axion fields. We analyze
Einstein-Rosen solutions of Einstein-Maxwell-dilaton-axion equations and
show, by explicitly taken the asymptotic limits, that they have
asymptotically velocity-term dominated (AVTD) singularities.
\end{abstract} 


\section{Introduction}
   
In the context of the so called string cosmologies there are several
aspects of interest. One of them is the way they approach to the
singularity as well as the asymptotic behaviour of spacetimes filled with
coupled scalar and U(1) fields.  Those matter fields arise naturally from
low energy effective superstring theory \cite{Shapere} i.e. the
Einstein-Maxwell-dilaton-axion (EMDA) system that contains the dilaton
which is non-minimally coupled with the Maxwell and axion fields. The EMDA
system has been discussed actively in the context of the black hole
solutions and singularities, and possesses characteristic features
different from the Einstein-Maxwell system \cite{MTN1}.

In this paper we investigate the character of initial cosmological
singularities in the presence of matter fields arising from the low energy
superstring theory.
 
Belinskii, Khalatnikov and Lifshitz (BKL) \cite{BKL} conjectured that the
dynamics of nearby observers would decouple near the singularity for
different spatial points. BKL also speculated that a generic singularity
should exhibit such a local oscilatory behavior. There is a special case
of the BKL conjecture called the {\it asymptotically
velocity-term-dominated} (AVTD) singularity that is not described by an
infinite sequence of Kasner epochs but by only one epoch. In this case no
oscillatory approach to the singularity is observed. The characteristic
feature of the AVTD singular behaviour is that all spatial derivative
terms of the field equations are negligible sufficiently close to the
singularity.
     
Another aspect of interest is to investigate the influence of nonminimal
exponential coupling of the dilaton and axion to the $U(1)$-field on the
character of the singularities. As it has been shown in \cite{BKL} and
\cite{berger} the minimally coupled scalar field can suppress the
oscilatory behaviour; on the other hand, it has been shown numerically
that magnetic Gowdy spacetimes are not AVTD but mixmaster \cite{berger2}.
It is therefore important to study the behavior of models containing both,
scalar and electromagnetic fields, in the limit $t \to 0$.

Our aim is to verify BKL conjecture in non-vacuum and spatially
inhomogeneous cases. To this end we shall consider Einstein-Rosen
spacetimes.

The Einstein-Rosen models are important among other things because many of
the non-perturbatively exact superstring backgrounds constructed from the
gauged Wess-Zumino-Witten (WZW) models admit two Abelian isometries
\cite{Tseytlin}. Hence, given our current knowledge of string cosmology
and conformal field theory, the $G_2$ Einstein-Rosen cosmologies derived
within the context of the low energy effective action represent a set of
models that is closely related to many exact string solutions in four
dimensions. Moreover, the Einstein-Rosen solutions include a number of
spatially homogeneous Bianchi models as special cases and certain $G_2$
models may be viewed as inhomogeneous generalizations of those Bianchi
cosmologies.
 
One subtle point in this treatment is that we are extrapolating asymptotic
behavior of solutions that are valid in the weak coupling regime to
regions near the big-bang singularity, for instance. In the strong
coupling regime one expects that higher-order corrections to the
perturbative theory, as well as non-perturbative string effects, should
become increasingly important. Hence, in this regime the qualitative
behavior of these solutions may deviate somewhat from that of solutions
derived from the full M or string theory. However, there are reasons to
believe that $G_2$ solutions should nevertheless provide a generic
description of cosmological models in the vicinity of a singularity. A
major incentive for this comes from the long standing conjecture of
Belinskii and Khalatnikov \cite{BKL}. This states that on the approach to
the cosmological singularity, the generalized Einstein-Rosen metrics may
play the role of the leading-order approximation to the general solution
of conventional Einstein gravity.
   
This paper is structured as follows: We present the EMDA field equations
and solutions in Secs. 2 and 3, respectively. In Sec. 3 we identify the
previous solution as an Einstein-Rosen spacetime and determine its
kinematical parameters. We analyze the Raychaudhuri equation to verify the
existence of singularity in Sec. 4. In Sec. 5 we perform the asymptotic
analysis of the metric and the fields, and show that the approach to the
singularity is of the form AVTD. In Sec. 6 we present the plane wave and
cylindrically symmetric cases and show their approaching to $t=0$.
Finally, some conclusions are given in Sec. 7.

\section{EMDA Einstein-Rosen solution}

As we pointed out above, we shall consider solutions to the low energy
effective theory for the heterotic string that arises as a dimensional
reduction and truncations of the string theory in four dimensions under
the following considerations: compactifications of six of the ten
dimensions and omission of the arising massless fields in the obtained
heterotic structure; only U(1) charges are permitted. The dynamical
equations of the resulting theory can be deduced from the action

\begin{equation}
\label{emda-action} 
S= \int d^4 x\sqrt{-g}
\{R-2(\partial\phi)^2-
\frac{e^\phi(\partial\kappa)^2}{2}-e^{-2\phi} F^2 -
\kappa F_{\mu\nu}\widetilde{F}^{\mu\nu}\},
\end{equation}
 
where $R$ is the scalar Riemann curvature, $g_{\mu \nu}$ is the
four-dimensional metric tensor, $F_{\mu \nu}$ is the electromagnetic
antisymmetric tensor field, $\widetilde{F}_{\mu\nu}$ its dual
($\widetilde{F}_{\mu\nu}=-\frac{1}{2}
\sqrt{-g}\epsilon_{\mu\nu\alpha\beta}F^{\alpha\beta}$), $\phi$ is the
dilaton scalar field and $\kappa$ is the axion field dual to the three
index antisymmetric tensor field $H=- \frac{1}{2}e^{4\phi}\ast d \kappa$.
The Einstein frame is related to the string frame by a conformal
transformation: $G_{\mu \nu}(s)=e^{2 \phi}g_{\mu \nu}(E)$, where $s$ and
$E$ stand for string and Einstein, respectively.

Previous work on these lines includes solution generating techniques which
permit to construct exact inhomogeneous solutions to the equations of low
energy string theory containing nonminimally coupled dilaton and
electromagnetic fields (EMD) presented in \cite{Yazadj}.  Solutions to
Einstein-dilaton-axion theory have been derived by applying the global
symmetries of the string effective action to a generalized Einstein-Rosen
metrics in \cite{Feinstein}. An algorithm for generating families of
inhomogeneous spacetimes with a massless scalar field was presented in
\cite{Lazkoz}.

In the Einstein-Maxwell-dilaton-axion (EMDA) theory, the dilaton $\phi$,
axion $\kappa$, gravitational $g_{\mu \nu}$ and electromagnetic Maxwell
$F_{\mu \nu}$ fields ought to fulfil the EMDA field equations derived from
(\ref{emda-action}):

\begin{eqnarray}
T_{\mu\nu}&=&R_{\mu\nu} -  \frac{1}{2} g_{\mu\nu} R , \\
T_{\mu\nu} &=& 2\partial_{\mu}\phi
\partial_{\nu}\phi +\frac{1}{2}e^{4\phi}\partial_{\mu}\kappa
\partial_{\nu}\kappa \nonumber\\
&& + 2e^{-2\phi} [F_{\mu\sigma}
F_{\nu}^{\sigma}-\frac{1}{4}g_{\mu\nu}F^{2}] \nonumber\\
&&-g_{\mu\nu}[\partial^{\alpha}\phi \partial_{\alpha}\phi +
\frac{1}{4}e^{4\phi}\partial^{\alpha}\kappa
\partial_{\alpha}\kappa ],\\
0&=&\nabla_{\mu}\partial^{\mu}\phi - \frac{1}{2}e^{4\phi}
g^{\mu\nu}\partial_{\mu}\kappa \partial_{\nu}\kappa 
+ \frac{1}{2}e^{2\phi} F^{2}, \nonumber\\
F^{2}& :=& F_{\alpha\beta} F^{\alpha\beta}, \nonumber\\
0&=&\nabla_{\mu} (e^{4\phi}\partial^{\mu}\kappa) -
F_{\mu\nu}\tilde{F}^{\mu\nu}, \\
J^{\nu}& :=& \nabla_{\mu} ( \kappa \tilde{F}^{\mu\nu} +
e^{2\phi}F^{\mu\nu})
\end{eqnarray}

In the next section we present EMDA solutions for metrics that possess two
spacelike Killing vectors. The fundamental structural functions of these
solutions are rational functions expressible as the ratio of polynomials
of at most second degree in the coordinate variables. These solutions were
obtained by straightforward integration of Eqs. (2)-(5) in \cite{breton}.

\subsection{Metric functions, fields and Weyl scalars}
 
The metric, endowed with two Killing vectors: $\partial_x$ and
$\partial_y$, is proposed in the form:
 
\begin{equation}
\label{metric(p,q)} 
ds^2 = \frac{\Delta}{P}dp^2 - \frac{\Delta}{Q}dq^2 +
\frac{P}{\Delta}(dx + Ndy)^2 + \frac{Q}{\Delta}(dx +Mdy)^2,
\end{equation}
 
The solutions are given by

\begin{eqnarray} 
P& =& \epsilon p^2 + 2np + \alpha, \nonumber\\
Q &=& \epsilon q^2+ 2mq - \alpha, \nonumber\\
M &= &\nu p^2 +2bp, \nonumber\\
N &=& -\nu q^2 + 2\beta q, 
\end{eqnarray}
together with $\epsilon = 1,0,-1$, $\nu = 1,0,-1$, $\Delta= M - N$. Being
$m, n, b, \beta, \alpha$ constants. There are distinct families of
solutions depending on the value of the kinematical constant $\epsilon =
1,0,-1$. The corresponding matter fields are given by

\begin{eqnarray}
e^{2\phi}(p,q) &=& \frac{\omega(p^{2}+q^{2})}{\Delta}, \\
\kappa(p,q) &=& \frac{2(bq+\beta p)}{\omega (p^{2}+q^{2})} + \kappa_0, \\
A_{x}(p,q) &=& -\frac{q_{0}q - g_{0}p}{\Delta}, \\
A_{y}(p,q) &=& \frac{\nu pq(q_{0}p + g_{0}q)}{\Delta},
\end{eqnarray}
where $\kappa_0$ and $\omega$ are axion and dilaton parameters,
respectively; while $A_{\mu}$ is
the electromagnetic potential and $g_0$ and $q_0$ are constants
related to the electromagnetic field. To fulfill the EMDA equations, the
constants $m, n, b, \beta, \alpha, g_0, q_0, \kappa_0, \omega, \epsilon,
\nu$ are restricted to satisfy the following algebraic conditions:

\begin{eqnarray} 
\nu^{2} q^{2}_{0}& =& 2\omega \beta (\beta \epsilon +m \nu), \quad
g_{0} \beta - q_{0}b = 0, \\
\nu^{2} g^{2}_{0} &=& 2\omega b (b\epsilon -n \nu), \quad
n \beta + mb = 0,
\end{eqnarray} 
  
This constrictions arise in the process of solving the field equations
with the metric functions proposed as polynomials. As a result, not all
the constants can be considered as free parameters; from the eleven
constants only five of them are free parameters.
  
The nonvanishing Weyl scalars are given by;

\begin{eqnarray}
\psi_{1} &=& \psi_{3} = -i\frac{\sqrt{PQ}}{2\Delta^{3}}(b^{2}+\beta^{2}),
\\
6\Delta^{3}\psi_{2}&=&-\{6\nu(n\nu - \epsilon b)
[p(p^{2}-3q^{2})+iq(q^{2}-3p^{2})]\nonumber\\
&&-6\nu(m\nu + \epsilon
\beta)[q(q^{2}-3p^{2})-ip(p^{2}-3q^{2})]\nonumber\\
&&+2[2 \epsilon(b^{2}+\beta^{2})+3\nu(m\beta-nb)](q^{2}-p^{2})\nonumber\\
&&+12ipq[(b\epsilon-n\nu)b+( \nu m + \epsilon
\beta)\beta]\nonumber\\
&&+4(nb-m\beta)(bp+\beta q)+4\alpha(\beta^{2}+b^{2})\},
\end{eqnarray}
   
Since the three Weyl scalars $\psi_{1}, \psi_{2}, \psi_{3}$ are nonzero,
the solution is of type G in the Petrov classification. From the above
expressions it can be seen that it is the presence of the constants $b$
and $\beta$ (non-diagonal terms in (\ref{metric(p,q)})) that gives such
structure, otherwise being a type D metric without axionic nor dilatonic
fields; these fields become constant when $b=0=\beta$. The invariants that
can be constructed from the Weyl scalars are $C^{(2)}= 6 (\psi_{2})^2-8
\psi_{1} \psi_{3}$ and $C^{(3)}=12 \psi_{2} \psi_{1} \psi_{3}-6
(\psi_{2})^3$. It can also be noted that such invariants will diverge
provided the Weyl scalars do.  This is the case if the function
$\Delta=M-N$ is zero, then essential singularities arise at those points.

\section{Einstein-Rosen spacetimes}

The line element (\ref{metric(p,q)}) can be transformed into the
generalized Einstein-Rosen form:

\begin{equation}
\label{ER} 
ds^2 = e^{f} (dz^2 - dt^2) + \gamma_{ab} dx^a dx^b.
\end{equation}
where $x^a=x,y$. All components are independent of the spatial coordinates
$(x,y)$; $f=f(z,t)$ determines the longitudinal part of the gravitational
field. The metric of the surfaces of transitivity is $\gamma_{ab}$ and the
gradient $K_{\mu}=\partial_{\mu} (det \gamma_{ab})^{1/2}$ determines the
local behaviour of the model. If $K_{\mu}$ is globally spacelike or null,
the solutions represent cylindrical and plane gravitational waves,
respectively. The cosmological models are characterized by a timelike
$K_{\mu}$ or when the sign of $K_{\mu}K^{\mu}$ changes.
 
In terms of the metric functions $\Delta$, $P$ and $Q$ of
(\ref{metric(p,q)}) the Einstein-Rosen metric is

\begin{equation}
\label{ERmetric}
ds^2 = \Delta(dz^2 - dt^2) + \frac{G}{\Delta}\{\chi (dx + Ndy)^2 
+ \chi^{-1} (dx+ M dy)^2\}
\end{equation}
 
where $\chi = \sqrt{P/Q}$ and $G= \sqrt{PQ}$. The appropriate coordinate
transformation $(p,q) \mapsto (z,t)$ that leads from (\ref{metric(p,q)})  
to (\ref{ERmetric})  depends on the values of the constants $\epsilon, n,
m, \alpha$. In Table \ref{table1} some cases are shown.

\begin{table}[h]
\begin{center}
\begin{tabular}{cccc}
\hline
$\epsilon$& p & q   & conditions \\\hline
 1   & $\sqrt{\alpha - n^2}\sinh{z} - n$  & $\sqrt{\alpha + m^2}\cosh{t}
- m$ & $\alpha - n^2 > 0 \;\;\;\ \alpha + m^2 > 0$ \\\medskip 
1 & ${e^{z}}/{2} - n$  & ${e^{t}}/{2} - m$ & $\alpha = n^2
= -m^2 = 0$ \\\medskip
 -1  & $\sqrt{\alpha + n^2}\sin{z} + n$ & $\sqrt{m^2 - \alpha}\sin{t}
+ m$ & $\alpha + n^2 > 0 \;\;\;\ m^2 - \alpha > 0 $ \\\medskip
 1 & ${e^{z}}/{2} - n$   & $\sqrt{\alpha + m^2}\cosh{t}
- m$  & $\alpha = n^2$  \\\medskip
 1   &$\sqrt{n^2 + m^2}\cosh{z}
- n$ & ${e^{t}}/{2} - m$ & $\alpha = - m^2$ \\\medskip
 0  & $\frac{z^{2}n^{2}-\alpha}{2n}$ & $\frac{t^{2}+\alpha}{2m}$ &  $-$
\\\hline
\end{tabular}
\end{center}
\caption{\small Coordinate transformations $(p,q) \rightarrow (z,t)$
to pass from the line element (\ref{metric(p,q)})
to the Einstein-Rosen form (\ref{ERmetric}).}
\label{table1}
\end{table}

The calculus of the gradient of the transitivity surface shows that the
three possibilities of interpreting the solution (\ref{ERmetric}) are
available as particular cases. We are interested in particular in the
cosmological model obtained with the transformation $(p,q) \mapsto (z,t)$
given by

\begin{eqnarray}
p &&\mapsto \sqrt{\alpha -n^2} \sinh{z}-n, \nonumber\\
q &&\mapsto \sqrt{\alpha +m^2} \cosh{t}-m. 
\end{eqnarray}

This transformation in the line element (\ref{ERmetric})  gives
 
\begin{eqnarray}
\Delta ds^2&&= \Delta^2(dz^2-dt^2)+(\alpha -n^2) \cosh^2{z}(dx+N dy)^2
\nonumber\\
&&+(\alpha +m^2) \sinh^2{t}(dx+M dy)^2,
\label{cosmetric}
\end{eqnarray}
where $\Delta =M(z)-N(t)$ is a monotonically increasing function on $z$
and $t$, given by

\begin{eqnarray}
\Delta&&=\nu(\sqrt{\alpha-n^2} \sinh{z}-n)^2 
+2b(\sqrt{\alpha-n^2} \sinh{z}-n) \nonumber\\
&&+ \nu(\sqrt{\alpha+m^2} \cosh{t}-m)^2
-2\beta(\sqrt{\alpha+m^2} \cosh{t}-m).
\end{eqnarray}
  
In our case $\Delta \ne 0$ over the whole spacetime and it implies that
the invariants are regular as well. The fact that the invariants are
regular and that the spacetime has a nonvanishing acceleration, as it is
shown in the next section, might be an indicative of absence of
singularity. A nonvanishing acceleration may create a gradient of pressure
which acts opposing gravitational attraction and this fact allows to avoid
the focusing of the congruence in the Raychaudhuri equation. A
singularity-free spacetime due to its acceleration was given in
\cite{Seno}. In the next subsection we characterize the cosmological model
using kinematical parameters.

\subsection{Kinematical parameters}
 
The kinematical characteristics of the Einstein-Rosen spacetime
(\ref{ERmetric}) are determined with respect to the timelike vector $u^a =
\frac{1}{\sqrt{\Delta}}\delta^a_t$, $u^{a}u_{a} = -1$. The acceleration,
expansion, deceleration and shear are given, respectively, by:

\begin{eqnarray}
\label{kinematics}  
\dot{u}^a _{z}&=& \frac{M^{'}}{2\Delta^2}, \nonumber\\ 
\Theta &=& \frac{\dot{Q} \Delta -\dot{N}Q}{2\Delta^{3/2}Q}, \nonumber\\
q &=& \frac{1}{(\dot{Q}\Delta -\dot{N}Q)^{2}}
\{8\dot{N}^2Q^2+5 \Delta^2 \dot{Q}^2 \nonumber\\
&&-\Delta Q \dot{N}\dot{Q} +6 \Delta Q (Q \ddot{N} - \Delta \ddot{Q}) \},
\nonumber\\
6 \sigma^2 &=& \frac{\dot{Q}^2}{ \Delta Q^2}+ \frac{4
\dot{Q}\dot{N}}{\Delta^2 Q} 
+ \frac{\dot{N}^2 (3P+4Q)}{\Delta^3 Q}, 
\end{eqnarray}
being null the rotation or vorticity, $\omega_{ab}=0$. For the
cosmological model the kinematical parameters behave asymptotically as
follows

\begin{eqnarray}
\Theta(t \to 0) &\to& \infty, \quad \Theta(t \to \infty) \to 0. \\
\sigma(t \to 0) &\to& \infty, \quad \sigma(t \to \infty) \to 0. \\
q(t \to 0) &\to& {\rm const}, \quad q(t \to \infty) \to {\rm const}.
\end{eqnarray}
    
\begin{figure}\centering
\epsfig{file=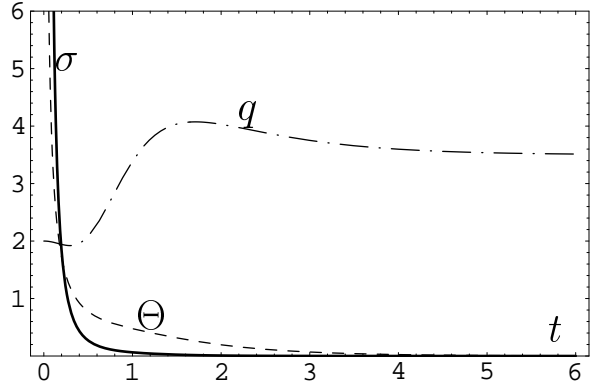, width=8cm}
\caption{Illustration of the behaviour of kinematical parameters for the
cosmological solution. Deceleration parameter, $q$, shear, $\sigma$, and
the expansion, $\Theta$, are shwown.  For this plot the constants have
been fixed to $m=n=2$, $b= -\beta=1$, $\alpha=5$ and $\nu=1$.}
\label{kin_par}
\end{figure}

This behaviour is illustrated in Fig \ref{kin_par}.  The shear evolves
towards isotropy for large times, $\sigma^2 \to 0$. The expansion, being
infinite at the origin of time, it decreases as time passes. The
deceleration is positive all the time and tends to be constant, this
indicates that the spacetime expands in non accelerated way. Both, shear
and expansion diverge at $t=0$, indicating the presence of a singularity;
the analysis of the Raychaudhuri equation in the next section confirms
that the spacetime has a spacelike singularity at $t=0$.

\section{Raychaudhuri equation}

According to Raychaudhuri, the equation that governs the rate of change of
expansion of timelike congruences, $\Theta$, is

\begin{equation}   
\frac{d \Theta}{d \lambda}= -R_{ab}u^{a}u^{b}+2 \omega^2-2
\sigma^2-\frac{\Theta^2}{3}+\dot {u}^{\alpha}_{; \alpha}
\label{Ray}
\end{equation}  
where $u^{\alpha}$ is a tangent vector and $\lambda$ is the affine
parameter of the congruence, $R_{ab}$ is the Ricci tensor, $\omega$ is the
rotation or vorticity, $\sigma$ is the shear and $\dot {u}^{\alpha}$ is
the acceleration of the congruence.

From Eq. (\ref{Ray}) it can be seen that the expansion $\Theta$ of a
timelike geodesic congruence with zero vorticity will monotonically
decrease along a geodesic if, for any timelike vector $V^a$,
$R_{ab}V^{a}V^{b} \ge 0$, i.e. if the strong energy condition (SEC) is
hold. The term involving the Ricci tensor, $R_{ab}$, in Eq. (\ref{Ray})
induces contraction of the geodesic lines, indicating that the focusing of
neighbouring geodesics is unavoidable if SEC is fulfilled since the other
terms on the right hand side of (\ref{Ray}) are also negative, with
exception of the acceleration; however, as we shall see the positive term
of the acceleration is defeated by the term $R_{ab} u^a u^b$.  
Substituting from ( \ref{kinematics}) in Eq. (\ref{Ray}) we obtain:

\begin{eqnarray}
\label{Raych}
R_{ab} u^a u^b &&=  \frac{1}{4 Q^2 \Delta}(\dot{Q}^2 - 2Q \ddot{Q})
-\frac{1}{2 \Delta ^3}(M^{'2}+\frac{P}{Q}\dot{N}^2) \nonumber\\
&& +\frac{1}{4\Delta^2}( M'
\frac{P{'}}{P}-3\dot{N}\frac{\dot{Q}}{Q}+
2M^{''}+2\ddot{N}), 
\end{eqnarray}
 
Moreover, from the expression for $\dot{\Theta}= d \Theta/ d \lambda$, Eq.
(\ref{Ray}), it can be seen that in fact it diverges as $t$ approaches the
origin provided the metric function $Q$ be divergent,

\begin{equation}
\label{dot-Th}
\dot{\Theta}=
\frac{1}{2\Delta}\{ \frac{\ddot{Q}}{Q}-\frac{\dot{Q}^2}{Q^2}
+\frac{\dot{Q}\dot{N}}{2Q \Delta}-\frac{3\dot{N}^2}{2\Delta^2}
- \frac{\ddot{N}}{\Delta}\}.
\end{equation}

In the cosmological case, $Q= (\alpha +m^2) \sinh^2{t}$, therefore as $t
\to 0$, $\dot{\Theta} \approx -1/\sinh^2{t} \to - \infty$, revealing the
existence of a singularity as $t \to 0$. This behaviour can be confirmed
by writting the derivatives of the coordinates with respect to an affine
parameter and checking that their coefficients diverge. By showing that
the first derivatives are unbounded one concludes that the corresponding
geodesic curves ${t}(\tau)$, ${x}(\tau)$, ${y}(\tau)$, ${z}(\tau)$, are
uncomplete at $t=0$ \cite{Arnold}.

\section{Asymptotic Behaviour}

In the limit $t \to 0$ the fields become functions that depend on $z$ but
otherwise are constant,

\begin{eqnarray}
e^{2\phi}(t \to 0)  && \rightarrow \frac{\omega
(A^2+(\sqrt{\alpha+m^2}-m)^2)}{\Delta},
\nonumber\\
\kappa(t\to 0) && \rightarrow \frac{2b(\sqrt{\alpha+m^2}-m)+2 \beta
A}{\omega 
(A^2+(\sqrt{\alpha+m^2}-m)^2)}+\kappa_{0}, \nonumber\\
A_{x}(t\to 0)&& \rightarrow \frac{g_0A-q_0(\sqrt{\alpha+m^2}-m)}{\Delta},
\nonumber\\
A_{y}(t\to 0)&& \rightarrow \frac{\nu
A(\sqrt{\alpha+m^2}-m)(q_0A+g_0(\sqrt{\alpha+m^2}-m))}{\Delta},
\end{eqnarray}

In relation to the asymptotic behaviour as $t \to \infty$, both scalar
fields, axion and dilaton tend to constant values as $t$ approaches
infinity, therefore the fields decouple for large times,

\begin{eqnarray}
\phi(t \to \infty)  && \rightarrow \frac{1}{2}\ln\frac{\omega}{\nu},
\nonumber\\ 
\kappa(t\to \infty) && \rightarrow \kappa_{0}, \nonumber\\
A_{x}(t\to \infty)&& \rightarrow 0, \nonumber\\ 
A_{y}(t\to \infty)&& \rightarrow Ag_{0},
\end{eqnarray}

There exists a lower bound on the value of the dilaton field and this
implies the existence of a lower, non-vanishing bound on the string
coupling which, in the context of M theory, in turn implies the existence
of a lower bound on the radius of the eleventh dimension \cite{witten}.
Moreover, in the limits that the volume of the transverse space measured
in the string frame becomes vanishingly small or arbitrarily large the
axion field is constant, since $\Gamma=$det$\Gamma_{ab} = e^{\phi} PQ$,

\begin{eqnarray}
\lim_{t \to 0} \Gamma&=&(\alpha-n^2)(\alpha+m^2)(\cosh^2{z}) t^2 \to 0,
\qquad
\kappa \to {\rm const}, \nonumber\\
\lim_{t \to \infty}
\Gamma&=&\frac{\omega}{4\nu}(\alpha-n^2)(\alpha+m^2)(\cosh^2{z}) e^{2t}
\to \infty, \qquad  \kappa \to \kappa_0.
\end{eqnarray}

Thus the two-form potential effectively decouples from the field equations
in these limits.

As was mentioned at the begining, one of the most interesting aspects of
analyzing cosmological models in to dilucidate how the spacetime behaves
near the initial singularity and to determine if this is an oscilatory
approach or a AVTD behaviour.

Narita, Torii and Maeda \cite{NTM} studied the influence of the
exponential dilaton-electromagnetic coupling on the character of initial
singularities. They considered EMDA system with a $T^3$ Gowdy cosmology
and using the Fushian algorithm they showed that those spacetimes have in
general asymptotic velocity-term dominated singularities. Their results
mean that the exponential coupling of the dilaton to the Maxwell field
does not change the nature of the singularity. In what follows we arrive
to the same conclusion for the Einstein-Rosen spaces by taking directly
the limit $t \to 0$ in the metric (\ref{cosmetric}).

Taking the limit $t \to 0$ with $z=$constant, in the metric functions of
(\ref{cosmetric}) for the studied case we obtain the line element

\begin{equation}
\label{asympt-metric}
ds^{2} = k (dz^{2}-dt^{2})+ \frac{P(z)}{k}(dx +
Ndy)^2 +\frac{(\alpha+m^{2})t^{2}}{k}(dx + Mdy)^2,
\end{equation} 
where $k=$const. and the metric functions are ( remind that z has been
fixed to constant)

\begin{eqnarray}
P(z)&=&(\alpha-n^2) \cosh^2(z), \nonumber\\
M(z)&=&\nu [\sqrt{\alpha-n^2} \sinh(z)-n]^2+2b[\sqrt{\alpha-n^2}
\sinh(z)-n], \nonumber\\
N&=& - \nu [\sqrt{\alpha+m^2} -m]^2+2\beta[\sqrt{\alpha+m^2} -m].
\end{eqnarray}
 
The line element (\ref{asympt-metric}) corresponds to a Kasner metric with
$p_1=p_3=0$ and $p_2=1$. This establishes the AVTD behaviour of this
model, extending the conclusion by Narita {\it et al} to Einstein-Rosen
spaces.

 
\section{Plane-symmetric Wave and Cylindrical spacetime}
  
To establish a comparison of the behavior near the singularity, we take
the limit as $t \to 0$ in the plane-symmetric wave spacetime and the
cylindrical one associated to the metric (\ref{ERmetric}).  As quoted in
Sec. 3 the local behavior of the Einstein-Rosen spacetime has to do with
the character of the gradient of the transitivity surface. The spacetime
(\ref{ERmetric}) can be interpreted as a plane-symmetric wave space by
using the coordinate transformation

\begin{equation}
p \mapsto \exp{z}/2, \quad q \mapsto \exp{t}/2,
\end{equation}

then the metric (\ref{ERmetric}) takes the form

\begin{equation}
\label{metricplanewaves}
ds^2 = \Delta(dz^2 - dt^2) + \frac{e^{2z}}{4\Delta}(dx + Ndy)^2
+\frac{e^{2t}}{4\Delta}(dx+ M dy)^2.
\end{equation}
 
This metric behaves as a conformally plane spacetime when
$t \to 0$:

\begin{equation}
ds^2 = k(dz^2 - dt^2) + \frac{A^2}{4k}(dx +N_1dy)^2
+\frac{1}{4k}(dx+ M_1 dy)^2.
\end{equation}
 
where $A^2=\exp{2z}$, $M_1=\nu^2 \exp{2z}/4+ \beta \exp{z}$ are constants
for $z=$const. and $N_1=- \nu^2/4 +b$. The metric is completely regular at
$t \to 0$; besides, neither $\Theta$ nor $\sigma$ diverge at $t=0$, in
contrast with the cosmological case. Investigating the tendency of the
change in expansion along a congruence, $\dot \Theta$, we found that it is
finite for all the time range. However the possibility exists of a fine
tunning of the axion and dilaton parameters, $b$ and $\beta$, that make
$\Delta=0$ and in that case the spacetime becomes singular.

The cylindrically symmetric case arises if instead we make the coordinate
transformation,

\begin{equation}
p \mapsto \sqrt{m^2+n^2} \cosh{z}-n, \quad q \mapsto \exp{t}/2-m.
\end{equation}
   
In this case the norm of the gradient of the transitivity surface is
always positive, $G_aG^a= e^{2t}(m^2+n^2)/(4 \Delta)$ and the spacetime
can be interpreted as a cylindrical symmetric one; the form of the metric
is:

\begin{equation}
\label{metricylindric}
ds^2 = \Delta(dz^2 - dt^2) + \frac{(m^2+n^2) \sinh^2{z}}{\Delta}(dx +
Ndy)^2 +\frac{e^{2t}}{4\Delta}(dx+ M dy)^2.
\end{equation}

In this case the singularity at $z=0$ is timelike.


\section{Conclusions}
           
We studied AVTD cosmological Einstein-Rosen spacetimes with EMDA fields.
The matter fields show a tendency to decoupling for large times. Near the
singularity the dynamics at different spatial points decouples and the
metric has a spatially varying Kasner form. We conclude that the
nonminimally coupling of the dilaton and axion to the Maxwell field does
not change the nature of the singularity. In this sense we have extended
to generalized Einstein-Rosen spaces the result, founded in Gowdy spaces
by NTM, that exponential coupling of the scalar field does not necessarily
lead to Mixmaster behavior.
 
Further analysis on solutions with plane-wave and cylindrical symmetry
might be of interest as roughly is shown in Sec. 6. The asymptotics of the
fields resemble the cosmological case, however, each model approaches
$t=0$ in a very different way.





\begin{thebibliography}{99}

\bibitem{Shapere}
Shapere, A., Trivedi, S., Wilczek, F.: Dual Dilaton Dyons,  Mod. Phys.
Lett. {\bf A6} 2677 (1991).

\bibitem{MTN1}
Maeda, K., T. Torii, T., Narita, M.: Do naked singularities generically
occur in generalized thoeries of gravity?, Phys. Rev. Lett. {\bf
81} 5270 (1998).

\bibitem{BKL}
Belinskii, V., Khalatnikov, I.: Effect of scalar and vector fields on
the nature of the cosmological singularity,  Sov. Phys. JETP {\bf 36}
591, (1973).
Belinskii, V., Khalatnikov, I., Lifshitz, E.: A general solution of the
Einstein equations with a time singularity, Adv. Phys.{\bf
31} 639 (1982).

\bibitem{berger} 
Berger, B.: Influence of scalar fields on the approach to a
cosmological singularity, Phys. Rev. D {\bf 61} 023508-1, (1999).

\bibitem{berger2}
Weaver, M., Isenberg, J., Berger, B. K.: Mixmaster behavior in
inhomogeneous cosmological spacetimes, Phys. Rev. Lett. {\bf 80},
2984 (1998). Berger, B. K.: Hunting local mixmaster dynamics in spatially
inhomogeneous cosmologies,  Class. Quantum
Grav. {\bf 21} 581 (2004).


\bibitem{Tseytlin}
Tseytlin, A. A.: Exact solutions of closed string theory, Class. Quantum
Grav. {\bf 12} 2365 (1995).

\bibitem{Yazadj}
Yazadjiev, S. S.: Exact inhomogeneous Einstein-Maxwell-Dilaton
cosmologies, Phys. Rev. D {\bf 63} 063510, (2001).

\bibitem{Feinstein}
Clancy, D., Feinstein, A., Lidsey, J. E., Tavakol, R.:Inhomogeneous
Einstein-Rosen string cosmology, Phys. Rev D, {\bf
60}, 043503, (1999).

\bibitem{Lazkoz}
Lazkoz, R.: $G_1$ spacetimes with gravitational and scalar waves, 
Phys. Rev. D {\bf 60} 104008 (1999).

\bibitem{breton} 
Breton, N.: Exact solutions in Einstein-Maxwell-Dilaton-Axion Theory, in
{\it Recent Developments in Gravitation}, Proceedings of ERE99, Ed. by J.
Ib\'a\~nez, Univ. Pa\'{\i}s Vasco, Espa\~na (2000), 179-184.

\bibitem{Seno}
Chinea, F. J., Fern\'{a}ndez-Jambrina, L.,
Senovilla, J. M. M.: Singularity-free space-time, {Phys. Rev. D}
\textbf{45}, 481 (1992).

\bibitem{Arnold}
Arnold, V. I.: \textit{Ordinary Differential Equations},
(MIT Press, 1990).

\bibitem{witten}
Witten, E.: String theory dynamics in various dimensions,  Nucl. Phys.
{\bf B443}, 85 (1995).

\bibitem{NTM} 
Narita, M., Torii, T., Maeda, K.: Asymptotic singular behaviour of
Gowdy spacetimes in string theory,
Class. Quantum Grav. {\bf 17} 4597 (2000).
 

\end{thebibliography}
\end{document}